\documentclass[conference]{IEEEtran}
\IEEEoverridecommandlockouts
\usepackage{hyperref}
\usepackage{cite}
\usepackage{amsmath,amssymb,amsfonts}
\usepackage{algorithmic}
\usepackage{graphicx}
\usepackage{textcomp}
\usepackage{xcolor}
\usepackage{multirow}
\usepackage{arydshln}
\usepackage{booktabs} 
\def\BibTeX{{\rm B\kern-.05em{\sc i\kern-.025em b}\kern-.08em
    T\kern-.1667em\lower.7ex\hbox{E}\kern-.125emX}}

\newcommand{\y}{\mathbf{y}}
\newcommand{\z}{\mathbf{z}}

\newcommand{\medusab}{Medusa-Block~}
\newcommand{\medusah}{Medusa-Linear~}

\begin{document}


\title{Whisper in Medusa’s Ear: Multi-head Efficient Decoding for Transformer-based ASR}


\author{
\IEEEauthorblockN{Yael Segal-Feldman, Aviv Shamsian, Aviv Navon, Gill Hetz, Joseph Keshet}
\IEEEauthorblockA{\textit{aiOla Research, Israel}\\
yael.segal@aiola.com}
}

\maketitle

\begin{abstract} 
Large transformer-based models have significant potential for speech transcription and translation. Their self-attention mechanisms and parallel processing enable them to capture complex patterns and dependencies in audio sequences. However, this potential comes with challenges, as these large and computationally intensive models lead to slow inference speeds. Various optimization strategies have been proposed to improve performance, including efficient hardware utilization and algorithmic enhancements. In this paper, we introduce \emph{Whisper-Medusa}, a novel approach designed to enhance processing speed with minimal impact on Word Error Rate (WER). The proposed model extends the OpenAI's Whisper architecture by predicting multiple tokens per iteration, resulting in a 50\% reduction in latency. We showcase the effectiveness of Whisper-Medusa across different learning setups and datasets. 

\end{abstract}


\section{Introduction}
\label{sec:intro}

Transformer-based supervised models are currently achieving state-of-the-art automatic speech recognition (ASR) results. Specifically, \emph{OpenAI's Whisper} \cite{radford2023robust} is one of the most successful models and has approximately 1.5 billion parameters (in its largest version), which ensure high transcription accuracy. However, despite their impressive performance these large models tend to suffer from slow inference speeds.

Recently, several optimization strategies have been developed to enhance inference efficiency in transformers. These strategies encompass efficient hardware utilization, algorithmic improvements, and sophisticated pre-processing and post-processing methodologies. One of the most well-known solutions is \emph{Faster-Whisper} \cite{fasterwhisper2024}. This approach involves re-implementing OpenAI's Whisper model using \emph{CTranslate2} \cite{opennmt_ctranslate2_2024}, a high-performance inference engine optimized explicitly for transformer models. Another prevalent approach involves applying guided knowledge distillation and quantization to the Whisper model \cite{shao2023whisper, gandhi2023distil}, reducing the model's size and improving speed, at the cost of WER degradation.


\emph{Speculative decoding} \cite{leviathan2023fast, xia2024unlocking} represents an emerging family of techniques in natural language processing (NLP) designed to speed up inference in large models. By performing multiple decoding steps in parallel, it reduces the number of computationally expensive operations required while decoding.
Typically, a smaller and faster assistant model generates several candidate tokens, from which the most promising ones are selected, enhancing efficiency and effectiveness. The assistant model produces multiple candidate sequences, evaluated and scored by the counterpart large mode based on their likelihood or quality, with the best sequence chosen as the final output. Recently, various approaches for candidate evaluation have been proposed, each differing in methodology and implementation \cite{chen2023accelerating, miao2023specinfer, spectoraccelerating}.

However, for practitioners, developing an effective assistant model and maintaining both the original and the assistant models is challenging. To address this, self-assistance methods have been proposed, where the original model serves as its own assistant model. Studies like \cite{stern2018blockwise}, \cite{gloeckle2024better}, and \cite{cai2024medusa} train the original model to generate multiple predictions at each step, using different techniques to generate and evaluate candidate sequences. Other approaches \cite{liu2024kangaroo} and \cite{zhang2023draft} utilize only parts of the original model as the assistant model.

Inspired by Stern et al. \cite{stern2018blockwise} and Cai et al. \cite{cai2024medusa} in the NLP domain, our study extends speculative decoding methods for the speech domain. Instead of sequentially generating a single token at a time, we propose modifying the architecture of encoder-decoder transformer-based ASR to predict multiple tokens per decoding step. ASR transformer-based models work differently than most large language models (LLMs), implemented as decoder-only. In ASR, the transformer decoder processes the entire speech audio from the encoder, which might offer even better efficiency than the decoder-only architectures. We exemplify our approach to OpenAI's Whisper large model.

The contributions of this paper are as follows: (1) We propose a novel ASR approach, \emph{Whisper-Medusa}, utilizing Speculative Decoding; (2) We introduce two architectures to implement this approach; (3) We demonstrate the effectiveness and advantages of our method through a comprehensive evaluation of diverse multilingual benchmarks. Additionally, we release the open-source code for further research and development, available at: \textcolor{magenta}{\textit{\url{https://github.com/aiola-lab/whisper-medusa}}}.

\section{Method}
\label{sec:model}

\begin{figure*}[h]
    \centering
    \includegraphics[width=\linewidth]{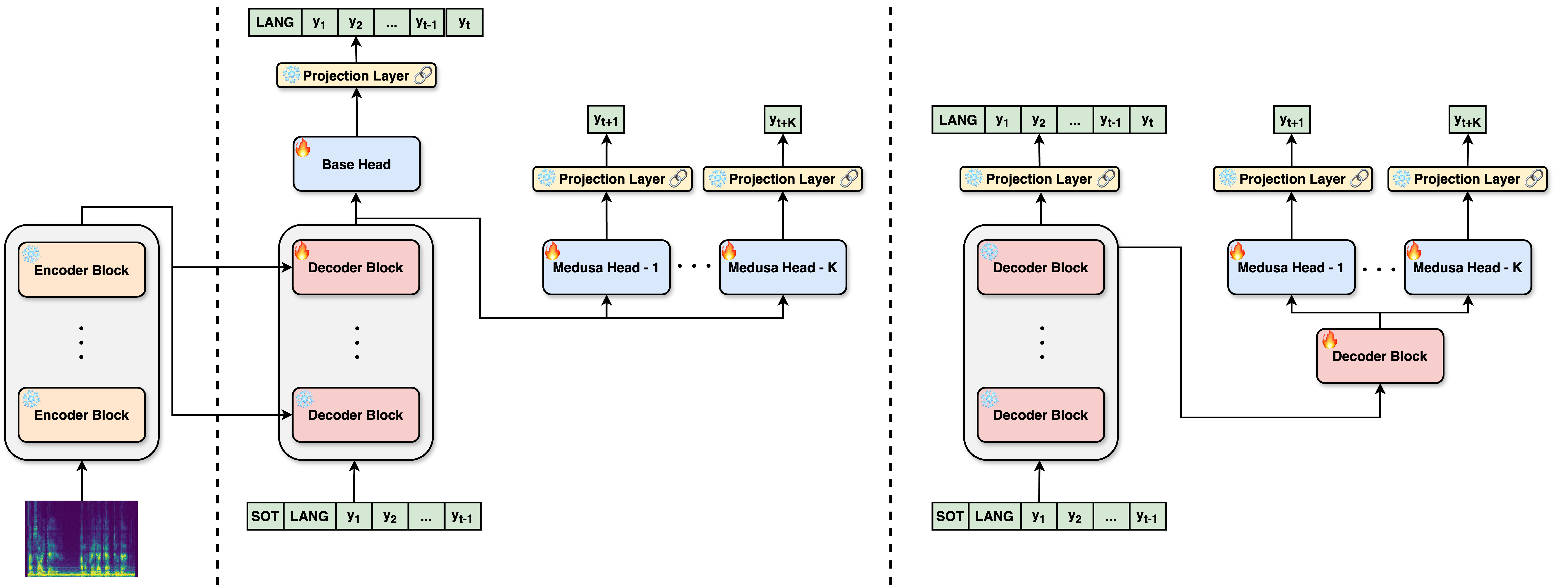}
    \caption{\textit{Whisper-Medusa Architectures}: Left - \textit{\medusah}: each Medusa head consists of a single linear layer with a residual connection, followed by a shared vocabulary projection layer (indicated by a chain symbol). Right - \textit{\medusab}: A Whisper decoder block shared across all Medusa heads, followed by a single linear layer and a residual connection for each head, with the outputs then passed to a shared vocabulary projection layer (indicated by a chain symbol).}
    \label{fig:medusa_architecture}
\end{figure*}

Transformer-based ASR architectures typically consist of an encoder and a decoder. The encoder processes the input audio waveform and converts it into a sequence of high-dimensional embeddings. Then, the decoder takes these embeddings as input and generates a sequence of tokens (usually sub-word units). The decoder operates autoregressively, predicting one token at a time. At each step, it estimates the probability distribution over the entire set of possible tokens and selects the most likely one. This process continues until an end-of-sequence token is generated.

Our focus is on modifying the decoder's behavior. Instead of predicting a single token at each iteration, we propose adjusting the decoder to simultaneously predict $K\!+1\!$ tokens. This approach aims to improve efficiency and potentially capture longer-range dependencies in the output sequence. In the following sections, we will rigorously describe the processes.

Let $y \in \mathcal{Y}$ denote a token from the set of tokens $\mathcal{Y}$. Denote by $\y_{<t}=(y_0, y_1, \ldots, y_{t-1})$ the sequence of tokens starting from the 0-th token $y_0$ until the $t\!-\!1$-th token, $y_{t-1}$. 

The decoder estimates the probability distribution of the next token $y_t$ given the predicted token sequence $\hat{y}_{<t}$ and the encoder's embeddings $\z$. That is, $p(y_t \mid \hat{\y}_{<t}, \z)$. Throughout the paper, we assume greedy decoding, where each token is predicted by selecting the one with the highest probability from the distribution:
\vspace{-0.5em}
\begin{equation}\label{eq:dec_pred}
\hat{y}_t=\underset{y_t}{\arg\max} \; p(y_t\mid \hat{\y}_{<t}, \z)~.
\end{equation}

We assume there are $K+1$ prediction heads. The \emph{base} head is the original decoder prediction head (before or after fine-tuning).  This base head generates the probability distribution $p_0$ over the set of tokens for the $t$-th token, conditioned on the preceding already predicted sequence of tokens $\hat{\y}_{<t}$, as described earlier. The probability distribution associated with the $k$-th head corresponds to the $y_{t+k}$ token, conditioned on the previous tokens $\hat{\y}_{<t}$, and is given by $p_k(y_{t+k} \mid \hat{\y}_{<t}, \z)$.

We describe the inference process using $K\!+\!1$ heads, assuming the model and its corresponding heads have already been trained. The inference includes two phases token prediction and verification. In the first phase, we estimate the distributions of all $K\!+\!1$ heads:
\vspace{-0.2em}
\begin{equation}
    p_0(y_t \mid \hat{\y}_{<t}, \z), ~ p_k(y_{t+k} \mid \hat{\y}_{<t}, \z) ~~ \text{for~} 1 \le k \le K ~,
\end{equation}
and identify the set of the subsequent $K\!+\!1$ tokens by selecting the tokens that maximize these probabilities: $\{\hat{y}_{t}, \hat{y}_{t+1}, \dots, \hat{y}_{t+K}\}$. 

In the second phase, we pass all the predicted tokens from the first phase through the base head and select all heads $0 \le k \le K$ for which the resulted probabilities are above a threshold:
\begin{equation}
    p_0(y_{t+k} \mid \hat{\y}_{<t}, \hat{y}_{t}, \dots, \hat{y}_{t+k-1}, \z) ~ > ~ \min\{\epsilon, \alpha \, \tilde{p}_{\max}\}~,
\end{equation}
where $\epsilon=0.09$ and $\alpha=0.3$ (selected on a held-out set), and $\tilde{p}_{\max}$ is the exponent of the entropy function\footnote{For a detailed discussion on the exponent of the entropy function see Leinster \cite{leinster2021entropy}.} of the distribution $p_0$. 

We introduce two architectures. The first, \medusah$\!\!$, features $K$ heads, each consisting of a single linear layer with a residual connection, followed by a \emph{shared} vocabulary projection layer to reduce parameters. We only update the final decoder layer and the heads to simplify training. We train the base head along with the additional $K$ heads by applying a cross-entropy (CE) loss to each one. The total loss is computed as the average of these individual losses, combined with a weighted KL-divergence loss between the probability distributions of the original ASR model (the weight was 0.01).
The second architecture, \medusab$\!\!$, includes an additional decoder block shared across all $K$ heads. This is followed by a single linear layer and a residual connection for each head, with the outputs then passed to a shared vocabulary projection layer. In this architecture, all the ASR model's weights are frozen, and only the Medusa weights are updated. Hence, we trained our model using $K$ weighted CE loss functions, where the base head is not trained, and without the KL loss function.
Our models are depicted in Fig.~\ref{fig:medusa_architecture}.

\begin{table*}
    \centering
    \caption{VoxPopuli: WER, CER  and speedup results for the Medusa models, Whisper Vanilla, and Whisper Fine-tuned, evaluated on the English, German, Spanish, and French subsets of Voxpopuli.}
    \resizebox{1\textwidth}{!}{%
    \begin{tabular}{lccccccccccccc} 
    \toprule
        &  & \multicolumn{3}{c}{English} & \multicolumn{3}{c}{German} & \multicolumn{3}{c}{Spanish} & \multicolumn{3}{c}{French}  \\

       Model& & WER $\downarrow$ & CER $\downarrow$ & Speedup& WER $\downarrow$ & CER $\downarrow$ &Speedup& WER $\downarrow$ & CER $\downarrow$ &Speedup& WER $\downarrow$ & CER $\downarrow$&Speedup \\
       \midrule
       Whisper Vanilla  &   &  8.40 & 5.03& - & 11.96 & 7.13 & -& \textbf{7.90}  &\textbf{ 5.23}  & -& 11.80 & 7.15 &- \\

       Whisper Fine-tuned  &   &  \textbf{6.51} & \textbf{4.08}&- &  \textbf{8.93} &  \textbf{5.01}& -&  9.96 &  6.43& - & \textbf{9.84} & \textbf{ 5.96} & -\\
        \medusah & &  9.91 & 6.96& \textbf{1.48}&  16.90 &  11.03& \textbf{1.31}&  14.22  & 8.75& \textbf{1.37}&16.29 &  11.22 & \textbf{1.40}\\
       \medusab& & 8.17 & 4.91& 1.40&11.77 & 6.77& 1.23 &8.20 &\textbf{5.23} &1.28 &11.87 &   7.07& 1.30 \\
       \bottomrule
    \end{tabular} %
    }
    \label{tab:vox4}
\end{table*}

\begin{table}
    \centering
    \caption{LibriSpeech: WER, CER and speedup results for the Medusa models, Whisper Vanilla and Whisper Fined-tuned, evaluated on the LibriSpeech Test sets.}
    \resizebox{0.5\textwidth}{!}{%
    \begin{tabular}{llccccc} 
    \toprule
       Dataset & Model & WER $\downarrow$ & CER $\downarrow$& Speedup $\uparrow$ & \# Params M\\
       \midrule
       \multirow{5}{*} {Test Clean}  & Whisper Vanilla &  4.00 & 1.86& - & 1550\\
         & Whisper Fine-tuned  & \textbf{2.23} & \textbf{0.74}& -& 1550\\
         & Whisper + Assistance\cite{gandhi2023distil}& 4.00 & 1.86& \textbf{1.57}& 1669\\
        & \medusah &  4.11 & 2.00 & 1.49& 1568\\
    
         & \medusab &  3.40& 1.42& 1.37& 1592\\
         \midrule
        \multirow{5}{*} {Test Other}  & Whisper Vanila & 6.19 & 2.96&-& 1550\\
         & Whisper Fine-tuned  & \textbf{4.66}& \textbf{1.97}&-& 1550\\
         & Whisper + Assistance\cite{gandhi2023distil}& 6.19 & 2.96& \textbf{1.49}&1669 \\
         & \medusah&  8.07& 4.21 & 1.40& 1568\\
          & \medusab &  6.16& 2.94 & 1.29&1592\\
         \bottomrule
    \end{tabular} %
    }
   \label{tab:libri}
\end{table}

\section{Experiments}
\label{sec:experiments}

\subsection{Datasets}
\label{sec:datasets}
We train and evaluate our models using two datasets.
First, we utilized LibriSpeech \cite{panayotov2015librispeech} dataset, a widely recognized resource in speech recognition research. LibriSpeech consists of approximately 1,000 hours of English read speech, sourced from public domain audiobooks, and is accompanied by corresponding transcriptions. Specifically, our model was trained on the LibriSpeech 100, 360, and 500 subsets. The original transcripts in LibriSpeech are in uppercase format, therefore we used the transcripts from \cite{meister2023librispeech}, which restores punctuation and capitalization for better readability. We evaluate our models using LibriSpeech's Test-Clean and Test-Other sets. Second, we used the VoxPopuli dataset \cite{wang-etal-2021-voxpopuli}, a large-scale multilingual speech corpus derived from European Parliament event recordings between 2009 and 2020. This dataset includes 400K hours of unlabelled speech data in 23 languages and 1,800 hours of transcribed speech data in 16 languages. We employ VoxPopuli in two experimental setups: fully-supervised and self-supervised learning.

\subsection{Experimental setup}
\label{sec:experimental_setup}
We integrated our Medusa heads into the Whisper Large-v2 model. The model was trained with a batch size of 16 for the LibriSpeech and fully-supervised Voxpopuli setups, and a batch size of 8 for the self-supervised setup of Voxpopuli, using a learning rate of 1e-4 and the Adafactor optimizer \cite{shazeer2018adafactor} across all setups. We trained the model for 200K/400K/500K update steps in LibriSpeech, fully supervised VoxPopuli, and self-supervised VoxPopuli setups, respectively. In all setups, we utilized early stopping on the validation set to select the best-performing model.


The input to our model follows a similar process to Whisper \cite{radford2023robust}, where audio is sampled at 16 kHz and converted into an 80-channel log-magnitude Mel spectrogram with a 25-millisecond window and a 10-millisecond stride. On inference, we applied an exponential decay length penalty re-scaling the model's logits. Specifically, we employ the regulation starting at the 140th token for the LibriSpeech dataset and at the 190th token for the VoxPopuli setups, using a regulation factor of 1.01 for all setups.

We compare our models to the original Whisper, a fine-tuned Whisper, and a speculative decoding method with an assistant model. The assistant model generates candidate tokens, which are validated by the main model in a single pass, retaining a token only if all previous tokens match the main model's output. Specifically, we used the model from \cite{gandhi2023distil}, which is available only for English, therefore we tested it solely on the LibriSpeech dataset. 
We did not apply an exponential decay length penalty to these models, as it did not enhance their performance. All speed evaluations were conducted on a single NVIDIA A10G GPU.

\section{Results}
\label{sec:results}

\subsection{Main results}
\label{sec:main_results}
We evaluated the speed performance and accuracy of \medusah and \medusab models on the Librispeech dataset, as shown in Table \ref{tab:libri}. The WER and CER for \medusab fall between those of Whisper vanilla and fine-tuned Whisper, leaning closer to Whisper vanilla due to its reliance on the un-tuned base Whisper head.
The \medusah model offers better speedup than \medusab but shows some WER and CER degradation due to fine-tuning. The Whisper with assistant model achieves the best speedup performance while maintaining similar results to Whisper vanilla, as dictated by its evaluation process. 
The \medusah adds 18M parameters, \medusab adds 42M, and the assistant model adds the largest amount of parameters, 119M.

Next, we proceeded to evaluate our model on the transcribed subset of the VoxPopuli dataset, focusing on English, German, French, and Spanish. We train our models simultaneously across all these languages, using 500 hours for English, 243 hours for German, 179 hours for French, and 132 hours for Spanish. Results are shown in Table 
\ref{tab:vox4}.
Again, the \medusab model demonstrates WER and CER results that lie between those of Whisper Vanilla and Whisper Fine-tuned, with speedup ranging from 1.40 for English, the highest, to 1.23 for German, the lowest. 
The \medusah model demonstrates better speedup results, with a peak of 1.48 for English and 1.31 for German, but achieves lower accuracy overall. This discrepancy may be due to the highly imbalanced training data, which makes the Medusa head training challenging.
In this context, it is notable that the performance of the fine-tuned Whisper model for Spanish is lower than the performance of the vanilla Whisper model.


\subsection{Self-supervised}
\label{sec:self_supervised}
\begin{table}
    \centering
    \small
    \caption{ Self-supervised VoxPopuli: WER, CER, and average speedup results for Medusa models and Whisper Vanilla. Evaluated on the Czech, Finnish, and Dutch, transcribed test sets.}
    \resizebox{0.5\textwidth}{!}{%
    \begin{tabular}{lcccc} 
    \toprule
       Subset & Model & WER $\downarrow$ & CER $\downarrow$ & Speedup  $\uparrow$\\
       \midrule
        \multirow{3}{*} {Voxpopuli Czech}   & Whisper Vanilla & 15.06 & 7.88 & -\\
          & \medusah& 14.70&  6.68& \textbf{1.83} \\
        & \medusab& \textbf{13.84} & \textbf{6.66}& 1.68\\
         \midrule
        \multirow{3}{*} {Voxpopuli Finnish}   & Whisper Vanilla &  \textbf{15.50} & 7.14&-\\
         & \medusah & 17.41&  \textbf{6.68}& \textbf{1.60}\\
         & \medusab& 15.94 & \textbf{6.68}& 1.46\\
          \midrule
     
        \multirow{3}{*} {Voxpopuli Dutch}   & Whisper Vanilla &\textbf{ 13.24}  &\textbf{ 8.13}&-\\
         & \medusah& 16.90&  9.87& \textbf{1.47}\\
         & \medusab & 14.04 & 8.41& 1.36\\
         \bottomrule
    \end{tabular} %
    }
    \label{tab:self_supervised}
\end{table}

We explored our model's performance in a self-supervised setup with the VoxPopuli dataset, which includes 400,000 hours of unlabeled speech. We selected three languages Czech, Finnish, and Dutch which transcribed them using Whisper-Large with a beam size of 5. 
We filtered out 5\% of examples with significant discrepancies between expected and actual transcript lengths and removed examples that appear in the test set of the transcribed VoxPopuli subset. 
This process resulted in 730 hours of transcribed data for Czech, 733 hours for Finnish, and 720 hours for Dutch. We trained both \medusab and \medusah on this data and evaluated the models on the corresponding languages from the transcribed VoxPopuli subset. Results are presented in Table \ref{tab:self_supervised}.
It can be seen that our models perform best in the Czech language. Notably, both \medusab and \medusah achieve higher accuracy than Whisper on Czech, despite being trained on Whisper's output.
In terms of speedup, our models achieve the best results for Czech and Finnish, with strong performance for Dutch as well. As in previous setups, \medusah delivers the best speed results and, unlike other setups, in this case, surpasses the Whisper vanilla model for both Czech and Finnish.
In this setup, both \medusah and \medusab perform least effectively for Dutch, the language with the fewest examples, showing lower WER and CER than the Whisper vanilla model and their lowest speed results. This behavior highlights the importance of data quantity and dataset balance for effective multi-language Medusa training.

\begin{figure}
    \vspace{-15pt}
    \centering
    \includegraphics[width=1\linewidth]{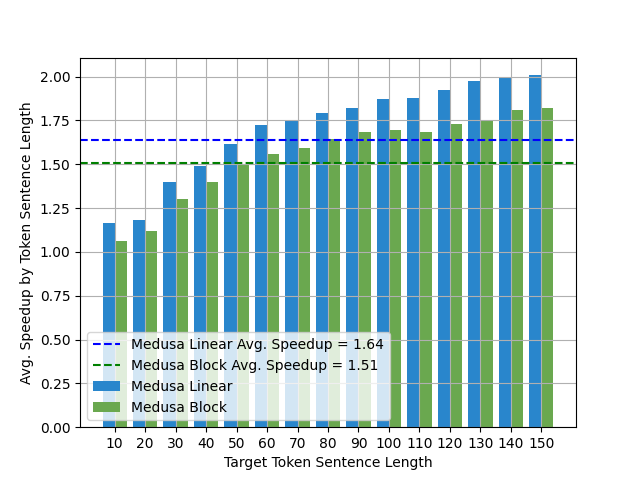}
  \caption{Average Speedup Results by Target Token Sequence Length for Czech, Finnish, and Dutch with the \medusah and \medusab Models.}
    \label{fig:speedup_by_length}
\end{figure}

Figure \ref{fig:speedup_by_length} presents the speedup results by target token sequence length for our self-supervised setup. 
The efficiency of the Medusa heads improves with longer target sequences, reaching a plateau of around 130 tokens. The Medusa heads are able to fully utilize their capabilities in longer sequences.

\subsection{Ablation}
\label{sec:ablation}

We evaluate the impact of the number of Medusa heads on the speed and accuracy of our \medusah model on the LibriSpeech dataset, as shown in Table \ref{tab:heads_ablation_start100}. 10 Medusa heads offer the best speed, while 5 heads achieve the best WER and CER. With 15 and 20 heads, both speed and accuracy decline.


\begin{table}
    \caption{ Number of Medusa Heads: WER, CER, and Average Speedup for the Base-head Medusa model on LibriSpeech Test sets, with varying Medusa heads.}
    \centering
    \small
    \begin{tabular}{lcccc} 
    \toprule
       Dataset & \# of heads & WER $\downarrow$ & CER $\downarrow$ & Speedup $\uparrow$ \\
       \midrule
        \multirow{4}{*} {Test Clean}  &  5 & \textbf{3.64} &\textbf{1.38}  & 1.42 \\
         & 10 &  4.11 & 2.00 & \textbf{1.49}\\
         & 15 & 5.74 & 3.41 &1.37\\
         & 20& 6.40& 4.21& 1.27\\
         \midrule
        \multirow{4}{*} {Test Other}  & 5&  7.00& 3.32 & 1.34\\
          & 10& 8.07& 4.21 &\textbf{1.40} \\
         & 15 & 9.21 & 6.86 & 1.30\\
         & 20 &10.87& 7.45 & 1.21\\
         \bottomrule
    \end{tabular} %
    \label{tab:heads_ablation_start100}
\end{table}

\subsection{Conclusions}
\label{sec:conclusions}
In this paper, we introduce \emph{Whisper-Medusa}, a novel ASR approach that leverages Speculative Decoding. To implement this approach, we present two architectures: \medusah and \medusab. Our findings demonstrate that \medusab achieves WER and CER results comparable to the original Whisper model, with latency reductions between 20\% to 68\%, depending on the dataset. On the other hand, \medusah delivers greater latency reductions, ranging from 30\% to 80\%, but with some performance degradation.

For future work, we aim to enhance the performance of \medusah, which is affected by imbalance  distribution of the dataset, and develop a specialized beam search designed for the Medusa method.

\bibliographystyle{IEEEbib}
\bibliography{refs}

\begin{thebibliography}{10}

\bibitem{radford2023robust}
Alec Radford, Jong~Wook Kim, Tao Xu, Greg Brockman, Christine McLeavey, and Ilya Sutskever,
\newblock ``Robust speech recognition via large-scale weak supervision,''
\newblock in {\em International conference on machine learning}. PMLR, 2023, pp. 28492--28518.

\bibitem{fasterwhisper2024}
SYSTRAN,
\newblock ``Faster whisper: High-performance whisper inference,'' https://github.com/SYSTRAN/faster-whisper, 2023.

\bibitem{opennmt_ctranslate2_2024}
{OpenNMT},
\newblock ``Ctranslate2,'' https://github.com/OpenNMT/CTranslate2, 2019.

\bibitem{shao2023whisper}
Hang Shao, Wei Wang, Bei Liu, Xun Gong, Haoyu Wang, and Yanmin Qian,
\newblock ``Whisper-kdq: A lightweight whisper via guided knowledge distillation and quantization for efficient asr,''
\newblock {\em arXiv preprint arXiv:2305.10788}, 2023.

\bibitem{gandhi2023distil}
Sanchit Gandhi, Patrick von Platen, and Alexander~M Rush,
\newblock ``Distil-whisper: Robust knowledge distillation via large-scale pseudo labelling,''
\newblock {\em arXiv preprint arXiv:2311.00430}, 2023.

\bibitem{leviathan2023fast}
Yaniv Leviathan, Matan Kalman, and Yossi Matias,
\newblock ``Fast inference from transformers via speculative decoding,''
\newblock in {\em International Conference on Machine Learning}. PMLR, 2023, pp. 19274--19286.

\bibitem{xia2024unlocking}
Heming Xia, Zhe Yang, Qingxiu Dong, Peiyi Wang, Yongqi Li, Tao Ge, Tianyu Liu, Wenjie Li, and Zhifang Sui,
\newblock ``Unlocking efficiency in large language model inference: A comprehensive survey of speculative decoding,''
\newblock {\em arXiv preprint arXiv:2401.07851}, 2024.

\bibitem{chen2023accelerating}
Charlie Chen, Sebastian Borgeaud, Geoffrey Irving, Jean-Baptiste Lespiau, Laurent Sifre, and John Jumper,
\newblock ``Accelerating large language model decoding with speculative sampling,''
\newblock {\em arXiv preprint arXiv:2302.01318}, 2023.

\bibitem{miao2023specinfer}
Xupeng Miao, Gabriele Oliaro, Zhihao Zhang, Xinhao Cheng, Zeyu Wang, Zhengxin Zhang, Rae Ying~Yee Wong, Alan Zhu, Lijie Yang, Xiaoxiang Shi, et~al.,
\newblock ``Specinfer: Accelerating generative large language model serving with tree-based speculative inference and verification,''
\newblock {\em arXiv preprint arXiv:2305.09781}, 2023.

\bibitem{spectoraccelerating}
Benjamin~Frederick Spector and Christopher Re,
\newblock ``Accelerating llm inference with staged speculative decoding,''
\newblock in {\em Workshop on Efficient Systems for Foundation Models@ ICML2023}.

\bibitem{stern2018blockwise}
Mitchell Stern, Noam Shazeer, and Jakob Uszkoreit,
\newblock ``Blockwise parallel decoding for deep autoregressive models,''
\newblock {\em Advances in Neural Information Processing Systems}, vol. 31, 2018.

\bibitem{gloeckle2024better}
Fabian Gloeckle, Badr~Youbi Idrissi, Baptiste Roziere, David Lopez-Paz, and Gabriel Synnaeve,
\newblock ``Better \& faster large language models via multi-token prediction,''
\newblock in {\em Forty-first International Conference on Machine Learning}, 2024.

\bibitem{cai2024medusa}
Tianle Cai, Yuhong Li, Zhengyang Geng, Hongwu Peng, Jason~D Lee, Deming Chen, and Tri Dao,
\newblock ``Medusa: Simple llm inference acceleration framework with multiple decoding heads,''
\newblock {\em arXiv preprint arXiv:2401.10774}, 2024.

\bibitem{liu2024kangaroo}
Fangcheng Liu, Yehui Tang, Zhenhua Liu, Yunsheng Ni, Kai Han, and Yunhe Wang,
\newblock ``Kangaroo: Lossless self-speculative decoding via double early exiting,''
\newblock {\em arXiv preprint arXiv:2404.18911}, 2024.

\bibitem{zhang2023draft}
Jun Zhang, Jue Wang, Huan Li, Lidan Shou, Ke~Chen, Gang Chen, and Sharad Mehrotra,
\newblock ``Draft \& verify: Lossless large language model acceleration via self-speculative decoding,''
\newblock {\em arXiv preprint arXiv:2309.08168}, 2023.

\bibitem{leinster2021entropy}
Tom Leinster,
\newblock {\em Entropy and diversity: the axiomatic approach},
\newblock Cambridge university press, 2021.

\bibitem{panayotov2015librispeech}
Vassil Panayotov, Guoguo Chen, Daniel Povey, and Sanjeev Khudanpur,
\newblock ``Librispeech: an asr corpus based on public domain audio books,''
\newblock in {\em 2015 IEEE international conference on acoustics, speech and signal processing (ICASSP)}. IEEE, 2015, pp. 5206--5210.

\bibitem{meister2023librispeech}
Aleksandr Meister, Matvei Novikov, Nikolay Karpov, Evelina Bakhturina, Vitaly Lavrukhin, and Boris Ginsburg,
\newblock ``Librispeech-pc: Benchmark for evaluation of punctuation and capitalization capabilities of end-to-end asr models,''
\newblock in {\em 2023 IEEE Automatic Speech Recognition and Understanding Workshop (ASRU)}. IEEE, 2023, pp. 1--7.

\bibitem{wang-etal-2021-voxpopuli}
Changhan Wang, Morgane Riviere, Ann Lee, Anne Wu, Chaitanya Talnikar, Daniel Haziza, Mary Williamson, Juan Pino, and Emmanuel Dupoux,
\newblock ``{V}ox{P}opuli: A large-scale multilingual speech corpus for representation learning, semi-supervised learning and interpretation,''
\newblock in {\em Annual Meeting of the Association for Computational Linguistics}. 2021, pp. 993--1003, Association for Computational Linguistics.

\bibitem{shazeer2018adafactor}
Noam Shazeer and Mitchell Stern,
\newblock ``Adafactor: Adaptive learning rates with sublinear memory cost,''
\newblock in {\em International Conference on Machine Learning}. PMLR, 2018, pp. 4596--4604.

\end{thebibliography}

\end{document}